\documentclass[pra, showpacs, notitlepage]{revtex4-1}

\usepackage{epsfig}
\usepackage{amssymb}
\usepackage{epstopdf}
\usepackage{amsmath}
\usepackage{bm}

\newcommand{\ket}[1]{{|{#1}\!\!>}}
\newcommand{\ip}[2]{{<\!\!{#1}|{#2}\!\!>}}

\begin{document}

\title{Lagrangian-Only Quantum Theory}
\author{K.B. Wharton}
\email{kenneth.wharton@sjsu.edu}
\affiliation{Department of Physics and Astronomy, San Jos\'{e} State University, San Jos\'{e}, CA 95192-0106}

\begin{abstract}
Despite the importance of the path integral, there have been relatively few attempts to look to the Lagrangian for a more realistic framework that might underlie quantum theory.  While such realism is not available for the standard path integral or quantum field theory, a promising alternative is to only consider field histories for which the Lagrangian density is always zero.  With this change, it appears possible to replace amplitudes with equally-weighted probabilities.  This paper demonstrates a proof-of-principle for this approach, using a toy Lagrangian that corresponds to an arbitrary spin state.  In this restricted framework one can derive both the Born rule and its limits of applicability.  The fact that the Lagrangian obeys future boundary constraints also results in the first continuous, spacetime-based, hidden-variable description of a Bell-inequality-violating system.

\end{abstract}


\maketitle

\setlength{\baselineskip}{1.2\baselineskip} 

\section{Introduction}

Feynman's discovery \cite{RPF} of a formal connection between the classical Lagrangian and canonical quantum mechanics (CQM) has been crucially important for theoretical physics, but has not led to much advancement in our understanding of CQM itself.  Well-known interpretational problems continue to present profound difficulties -- including the so-called ``measurement problem'', and the lack of ``realism''.  (``Realistic'' models are strictly defined here as those having a spacetime-representation of all events, even between measurements \cite{footnote0}.)  The few Lagrangian-based efforts to solve these problems have been strongly linked to the standard path integral.  By departing from the path integral, this paper develops a simple Lagrangian-based framework that defines a realistic model, out of which known quantum behavior seems to emerge (statistically) at a coarse-grained level.  The central proof-of-principle for this substantial claim is the derivation of the Born rule for spin states using a simplified Lagrangian.

One might argue that a Lagrangian framework cannot further quantum foundations research because of the formal equivalence between the predictions of CQM and the probabilities generated by the Feynman Path Integral (FPI).  This equivalence might be thought to justify an evident bias in the literature; most foundational research has been focused on CQM, as opposed to the FPI.  But there are several reasons to reconsider this focus.

First, there is a growing community that strongly argues quantum states are more naturally understood as states of knowledge rather than states of reality \cite{Fuchs, Spekkens, Harrigan}.  If this is indeed the case (and if the various ``no-go theorems'' can be bypassed, as discussed below), then the central question in quantum foundations is: \textit{What is the state of reality that underlies our knowledge $\psi$?}.  To the extent that the FPI describes a very different way to calculate the same probabilities, it also may indicate a different answer to this question, one worthy of consideration independent of the CQM formalism. 

A stronger argument is that many physicists have proposed modifications to CQM that formally break the equivalence to the FPI.  Adding small non-linear terms to the Schrodinger equation \cite{Pearle, GRW} or postulating non-equilibrium distributions of Bohmian particles \cite{Valentini} is worthwhile research in quantum foundations, and yet these approaches have no exact FPI analog.  It is similarly worthwhile to study modified-FPI approaches that have no exact CQM analog.  And yet, such modifications of the FPI are difficult to find in the literature.  This may represent a deep bias towards the initial-boundary/differential-equation framework as opposed to the spacetime-boundary framework implied by a literal reading of the FPI mathematics \cite{FQXi4}.

Indeed, a fresh assessment of the Hamiltonian vs. Lagrangian dichotomy might lead one to conclude that it is the Lagrangian that is likely more fundamental.  It is the Lagrangian (density) that most naturally unites space and time, and provides the natural parameterization-independence required by general relativity.  The Lagrangian also reveals the symmetries of the Standard Model.  The path from the Lagrangian to the Hamiltonian introduces the notorious ``problem of time'' in quantum gravity and ``appears to violate the very central lesson of relativity'' \cite{CW}.  The primary strength of the Hamiltonian formalism lies in quantum theory, and yet this is where the most unresolved foundational questions remain.  If it were possible to find a Lagrangian-only approach to quantum theory, the path to unification would be much more evident.  And as the following results indicate, a Lagrangian-based approach may comprise a realistic alternative to CQM.

The plan of the paper is as follows.  Section II details how close the path integral comes to implying a realistic underlying theory, and summarizes relevant prior research.  Section III introduces a (classical) toy Lagrangian for a spin state, which can be used as a simple test case in the subsequent analysis.  The central portion of the paper is the non-classical framework that is proposed as a replacement for the path integral (section IV), followed by applications of this framework to the toy Lagrangian (section V).  After deriving the limits in which the Born rule is recovered (and the limits in which one might expect deviations), section V.C develops a realistic account of entanglement.  Some deeper questions are addressed in the final section.

\section{Background}

\subsection{Realism and the Path Integral}

The path integral seems to imply that a particle takes all possible paths, which of course is not ``realistic''; as defined above, a realistic interpretation of a particle must hold that it took \text{one particular path}.  (Realistic fields need not be confined to a trajectory, but must have actual values as functions over spacetime, such as the classical electric field $\bm{E}(\bm{x},t)$.)   Still, one might imagine that a  ``one real path'' interpretation was nevertheless available, because each \textit{possible} path could be assigned a subjective probability.  Summing the probabilities of some class of paths could then yield the total probability for that class, even though only one path was actual.  In this case the probabilities would merely indicate our lack of knowledge of the actual path rather than a failure of realism.

Before discussing why such an interpretation is \textit{not} available for the standard FPI, it is worthwhile pointing out connections between such an approach and classical statistical mechanics.  The microcanonical ensemble is an example of how a realistic picture (the individual microstates) can underly a probabilistic configuration-space theory such as thermodynamics.  The goal of realistic approaches to CQM is to find an analogous underpinning to the quantum wavefunction.  The na\"ive story in the previous paragraph might be seen as a natural extension of statistical mechanics into spacetime, replacing 3D microstates with 4D ``micropaths'' (for particles), or ``microhistories'' (in general).  By associating each microhistory $\mu_j(\bm{x},t)$ with a (subjective) probability $p(\mu_j)$, states of knowledge are statistical ensembles that naturally live in a configuration space over the index $j$. 

In the most natural extension of statistical mechanics, one would simply find the probability of any set of constraints $C$ on $\mu_j$ according to 
\begin{equation}
\label{eq:JPD}
P(C)=\sum_{j_C} p(\mu_j),
\end{equation}
where $j_C$ are the microhistories compatible with the constraints $C$.  This is how the probability of 3D macrostates is found from the microcanonical distribution, the only real difference here being that the 4D boundary constraints $C$ include both preparation and outcome.  Note that since $C$ is traditionally divided up and treated as multiple constraints (preparation, outcome, spatial boundaries, etc.), $P(C)$ is most naturally viewed as a \textit{joint} probability distribution over all the constraints that make up $C$.  (While $P(C)$ is not a conditional probability, one can always use a complete set of joint probabilities to generate any related conditional probability.)  The key point is that Eqn (\ref{eq:JPD}) is a \textit{realistic} framework, in that the actual microhistory would in fact be one particular $\mu_{j'}$, and all probabilities would be attributable to a lack of knowledge of the actual $j'$.

Nevertheless, the FPI does not conform to the framework of Eqn (\ref{eq:JPD}).  Even in the specialized case of a single particle constrained by two consecutive position measurements, the FPI gives the (unnormalized) joint probability of those two measurements as
\begin{equation}
\label{eq:FPI}
P_{fpi}(C)=\left| \sum_{j_C} e^{iS(\mu_j)/\hbar} \right| ^2.
\end{equation}
Here the microhistory $\mu_j$ is evaluated for all particle paths $j_C$ that connect the two position measurements, $S$ is the classical action associated with that path, and the sum represents an integral over all such paths.

A comparison of (\ref{eq:JPD}) and (\ref{eq:FPI}) reveals several reasons why there is no realistic interpretation of the FPI.  For one, the summation is not over probabilities, but amplitudes that can be negative and even complex.  For another, squaring the sum in (\ref{eq:FPI}) looks dramatically different than the ordinary probability addition in (\ref{eq:JPD}).  Another relevant issue is that the $p$'s have an equal \textit{a priori} probability for each microstate in statistical mechanics, and this is evidently not the case for the amplitudes in (\ref{eq:FPI}).

When a particle is no longer exclusively constrained by \textit{position} measurements, the FPI gains new terms (see, \textit{e.g.}, \cite{WMP}) that make it even more difficult to compare to Eqn (\ref{eq:JPD}).  However, quantum field theory allows one to recover the original form (\ref{eq:FPI}), at the expense of replacing the (particle-based) path integral with a functional integral over all field configurations $\mu(\bm{x},t)$.  But quantum field theory comes no closer to Eqn (\ref{eq:JPD}) than does the FPI.   Furthermore, the allowed results of any given measurement cannot be deduced from the Lagrangian.  Instead, one must first determine the allowable outcomes using CQM before one can even use (\ref{eq:FPI}).  Along with issues involving the evaluation of the functional integral, this is one reason that neither FPI nor quantum field theory can be considered a ``Lagrangian-only theory''.  Still, these are certainly not the only barriers to a realistic framework along the lines of Eqn (\ref{eq:JPD}).

\subsection{Prior Work}

Although most quantum foundations research is based upon CQM, some Lagrangian-based approaches have been addressed in the literature.  Most notably, Sinha and Sorkin have demonstrated \cite{Sorkin} that it is possible to remove both the square and the imaginary portion of (\ref{eq:FPI}).  For particle-trajectories with well-defined endpoints, the most ``realistic'' reading of their mathematics is that each particle splits into two half-particles, and each half-particle takes an independent path between the endpoints.  The resulting microhistories $\mu_j$ each encode a pair of paths, and associated half-particle actions $S_1$ and $S_2$.  The joint probabilities from (\ref{eq:FPI}) can then be rewritten as
\begin{equation}
\label{eq:SaS}
P_{fpi}(C)= \sum_{j_C} cos\left( \frac{S_1(\mu_j)-S_2(\mu_j)}{\hbar} \right).
\end{equation}
This intriguing result clearly resembles the realistic framework (\ref{eq:JPD}), solving most of the problems mentioned above.  But as the authors note, the remaining problem is that the ``probabilities'' $p(\mu)$ can still be negative, which means that they are not probabilities at all.  In order to get interference in a particle-picture, the need for negative amplitudes seems to be inescapable.  It remains to be seen whether a field-picture (for which interference is more natural) might solve this last problem.

In another step towards realism, Griffiths' ``consistent histories'' framework \cite{Griffiths} notes that some between-measurement propositions do follow the rules of classical probability.  (For example, in a double interferometer, one can sometimes evolve the final measurement backward to deduce which arm the particle ``actually'' travelled through \cite{Hardy, WMP}, providing a fuller description than CQM permits.)  Unfortunately, this procedure is not universal; in the case of ``inconsistent histories'', no realistic account is available.  In terms of modifying the path integral itself, there has been some work in restricting the paths to a discrete lattice \cite{Stuckey}, with some interesting results \cite{Stuckey2}, but this has not led to an evident realistic underpinning to quantum theory.

Some progress has also been made on framing quantum measurement theory in a Lagrangian-only picture.  It is notable that the ``old quantum theory'' effectively utilized an action quantization to deduce energy eigenvalues, even before modern CQM was developed.  A semiclassical extension of this approach (EBK action quantization, with Maslov indices \cite{Maslov}) is further progress towards this end.  In addition, demanding strict action extremization (even on the boundary terms) effectively forces a quantization of measurement constraints for non-energy measurements such as angular momentum \cite{Wharton09}.  A full solution to this problem is beyond the scope of this paper, although there are some related developments in section IV.C.  Still, these partial results lend credence to the idea that a Lagrangian-only measurement theory is indeed possible.

Concerning what is ultimately possible, some may be under the impression that certain no-go theorems (namely Bell's theorem \cite{Bell} and the more recent PBR theorem \cite{PBR}) have already ruled out realistic interpretations along the lines of (\ref{eq:JPD}).  However, the scheme of assigning probabilities $p(\mu)$ to entire microhistories (rather than instantaneous states) provides a natural ``out'' from these theorems.  Crucially, the portion of a measurement constraint $C$ that occurs at time $t_f$ does not just constrain the instantaneous microhistory $\mu(\bm{x},t_f)$, but also the \textit{entire} microhistory, including times before $t_f$.  Different types of measurement could therefore entail different-sized possibility spaces $j_C$, and the actual microhistory is effectively constrained by future events.  This situation explicitly violates the ``outcome independence'' assumption \cite{Shimony} behind Bell's theorem, as well the ``preparation independence'' assumption behind the PBR theorem (if the systems are eventually brought together and jointly measured).  Such a conclusion is aligned with the mathematical duality between single-particle and entangled-particle joint probabilities \cite{Leifer,WMP,EPW,LnS}, indicating that any realistic Lagrangian-based interpretation for a single particle is also available for entangled particles.  An explicit proof-of-principle of these claims will be presented in section V.C, followed by further discussion.

Concerning the results below, the most important prior work is due to Schulman \cite{Schulman}.  This is a scenario where a standard quantum system is given seemingly-inexplicable ``kicks'' that just happen to knock it into one of the pure states in the measured basis.  In order to generate the correct relative probabilities, Schulman has shown that the ``kicks'' must be Cauchy-distributed, with a probability distribution over $\alpha$ (the rotation angle in Hilbert space) of $P(\alpha)\propto(\gamma^2+\alpha^2)^{-1}$.  From this distribution one can recover the Born rule in the limit $\gamma\to0$, but there are small deviations from the Born rule for a nonzero $\gamma$.  This result will be further discussed in section V.

\section{Spin States via Classical Fields}

\subsection{A Toy Lagrangian for Spin}

While any complete Lagrangian-only theory will require Lagrangian densities that span both space and time, the spin states of charged particles offer a simpler testbed for such approaches, in that extended physical space is no longer an absolute necessity.  It is possible to represent most of the essential physics of quantum spin in terms of field variables $q(t)$ that have no spatial extent, the standard CQM spin state $\ket{\chi(t)}$ being one example.  Spatial information still can be fed into $q(t)$ by virtue of a spatially-oriented magnetic field that influences $q(t)$ in the appropriate manner.  Furthermore, any oriented measurement (say, $\bm{S_x}$) can be accomplished by imposing a magnetic field in the corresponding direction (say, $\hat{x}$) and then making an energy measurement.  (Energy is well-defined without the need for a spatial extent of $q(t)$.)  Note that this is how spin angular momentum measurements are often made in the laboratory.

It is therefore possible to analyze quantum phenomena using a ``toy Lagrangian'' built from fields that are functions of time, but not space.  What gets lost from such a ``spaceless'' approach is the possibility of Larmor orbits or dipole-interactions with electric fields, but such effects are not needed for the basic quantum phenomena in greatest need of explanation.  Energy measurements of (charged) spins in controllable magnetic fields can be arranged to explore nearly every aspect of quantum phenomena, including Bell-inequality violations.  This allows a simpler proof-of-principle analysis without introducing the complication of a spatial extent for the fields.

Under this simplification, this toy theory uses a Lagrangian $L(q,\dot{q},t)$, not a Lagrangian density.  To keep the connection to CQM as evident as possible, if the traditional spin-state $\ket{\chi(t)}$ lives in an $n$-dimensional Hilbert space, define $\ket{q(t)}$ as another vector of $n$ complex numbers, and $\ket{\dot{q}}=(d/dt)\ket{q}$.  The ultimate goal is for  $\ket{q}$ to be a representation of what is actually happening, whether or not a measurement is made.  If it transpired that $\ket{q(t)}=\ket{\chi(t)}$, then the Copenhagen/Collapse interpretation of CQM would be recovered exactly (although this will not be the case).

Building from the results in \cite{WLSL}, consider the real Lagrangian defined by
\begin{eqnarray}
\label{eq:Lag}
L(q,\dot{q},t)=\ip{p}{p}-mc^2(\ip{q}{p}+\ip{p}{q}),
\\
\label{eq:pvec}
\ket{p}\equiv [mc^2-\gamma \bm{S}\cdot \bm{B}(t)]\ket{q}-i\hbar\ket{\dot{q}}.
\end{eqnarray}
Here $\gamma$ is the gyromagnetic ratio, $\bm{B}(t)$ is the external magnetic field, and $\bm{S}$ is the conventional vector of spin operators in an $n$-dimensional Hilbert space.  The vector $\ket{p}$ is nothing more than the definition (\ref{eq:pvec}), more familiarly written as $[\bm{H}-i\hbar \,d/dt]\ket{q}$, where $\bm{H}$ is the traditional Hamiltonian for a spin state plus the rest energy $mc^2$.  By definition, the traditional spin state $\ket{\chi}$ solves the Schr\"odinger-Pauli equation $[-\gamma\bm{S}\cdot\bm{B}(t)-i\hbar\, d/dt]\ket{\chi}=0$.  So if $\ket{q}$ is equal to $\ket{\chi}$ times a rest mass oscillation $exp(-imc^2t/\hbar)$, then $\ket{p}=0$.  And since global phases are irrelevant to $\ket{\chi}$, the usual Schr\"odinger-Pauli solutions also obey $\ket{p}=0$.  

\subsection{Classical Dynamics}

It is instructive to treat $L$ as a classical Lagrangian, and $q$ as a set of $2n$ classical (real-valued) parameters.  In this case, the Euler-Lagrange equations can be expressed as $2n$ coupled second-order differential equations, or combined into an easier-to-interpret expression:
\begin{equation}
\label{eq:ELE}
\left(mc^2+\gamma \bm{S}\cdot\bm{B}+i\hbar \frac{d}{dt}\right)\left(mc^2-\gamma \bm{S}\cdot\bm{B}-i\hbar \frac{d}{dt}\right)\ket{q(t)}=0.
\end{equation}
Notice that the rightmost `operator' on $\ket{q}$ yields $\ket{p}$.  The general solutions to these second-order equations are of the form
\begin{equation}
\label{eq:ELS}
\ket{q(t)}=\ket{q_+}+\ket{q_-},
\end{equation}
where $\ket{q_+}$ is any solution to $\ket{p}=0$ (\textit{i.e.} the traditional spin state) and $\ket{q_-}$ is any solution to $\ket{p}=2mc^2\ket{q}$. 

The solutions $\ket{q_-}$ bear some analogies to antimatter partners of $\ket{q_+}$.  Whether or not this is meaningful in such a restricted toy Lagrangian, one can imagine a broader theory in which some other measurement provided additional information concerning the initial system, restricting the solutions to either $\ket{q_+}$ \textit{or} $\ket{q_-}$.  This paper will primarily examine the restricted solution space $\ket{q(t)}=\ket{q_+}$ as a classical starting point.  Such solutions obey $\ket{p}=0$ and therefore exactly map to the Schr\"odinger-Pauli dynamics of an unmeasured spin state in an arbitrary magnetic field.  

This result may seem surprising, given that $\ket{q}$ is merely a representation of $2n$ coupled classical oscillators.  Nevertheless, this connection has been worked out explicitly for the case of spin-1/2, using the same Lagrangian as (\ref{eq:Lag}) despite different notation \cite{WLSL}.  The (seemingly) non-classical behavior of the unmeasured spin-1/2 system can nevertheless have a classical analog.  And of course, this is not a full recovery of CQM, because there is no restriction on measurement outcomes.  In other words, $\ket{q(t)}$ does not track the discontinuous collapse of $\ket{\chi(t)}$ upon measurement.

Finally, note that both of the solutions, $\ket{q_+}$ and $\ket{q_-}$, yield $L\!=\!0$ when plugged back into (\ref{eq:Lag}).  (Certain linear combinations of $\ket{q_+}$ and $\ket{q_-}$ also yield $L\!=\!0$.)  At this stage, this is merely a mathematical curiosity, on par with the fact that the solutions to the Dirac equation also yield ${\cal L}\!=\!0$ when used to calculate the Dirac Lagrangian density.  In section IV, however, this fact will be crucial.

\subsection{Hidden Parameters}

One difference between the classical $\ket{q}$ and the quantum spin state $\ket{\chi}$ is that $\ket{q}$ solves a second-order (in time) differential equation, while $\ket{\chi}$ solves the first-order Schr\"odinger-Pauli equation.  This explains why the solution space of $\ket{q}$ must be halved to match up with $\ket{\chi}$ (or, alternatively, why there is room for antimatter-like solutions in $\ket{q}$).  

However, even after halving the solution space there is still a slight mismatch.  The solutions to (\ref{eq:ELE}) can be exactly specified by initial conditions $\ket{q(0)}$ and $\ket{\dot{q}(0)}$, or $4n$ real parameters.  Halving this parameter space by restricting it to only $\ket{q_+}$ brings this down to $2n$ real parameters.  (The equation $\ket{p}=0$ is first-order in time.)  But the n-dimensional Hilbert space of $\ket{\chi}$ corresponds to $2n-2$ real parameters, leaving two ``hidden'' parameters in $\ket{q_+}$ that are not represented in the quantum state $\ket{\chi}$.

The first of these hidden parameters is the (so far) unconstrained amplitude $\ip{q_+}{q_+}$, as contrasted with the usual normalization $\ip{\chi}{\chi}=1$.  One charitable reading of this amplitude would be to relate it to particle number, in which case the initial condition corresponding to a single particle would fix the amplitude at some particular value.  Even without such a reading, $\ip{q_+}{q_+}$ is constant (classically), and is therefore determined by the initial values $\ket{q(0)}$.  Pending any physical interpretation of $\ip{q_+}{q_+}$, we can tentatively set this constant to 1.

The more-important hidden parameter is the global phase $\theta(t)$.  In CQM's projective Hilbert space this phase is irrelevant by design, as both $exp(i\theta)\ket{\chi}$ and $\ket{\chi}$ correspond to the same ray, but no such relation exists for the classical parameters $\ket{q(t)}$.  The global phase of coupled classical oscillators is clearly a physically meaningful quantity; one can distinguish oscillators with different phases but otherwise identical behavior.  And yet in quantum theory, the corresponding phase is typically thought to be an unphysical gauge.

Denying importance to $\theta(t)$ in quantum theory is all the more curious when one examines cases such as the two-particle Aharonov-Bohm effect \cite{2pAB}, and other relative phase measurements.  These are cases when the phase of a single particle \textit{is} empirically accessible.  The standard thinking is that (in these cases) one expands the Hilbert space to include both particles, and the meaningless global phase of a single particle becomes the meaningful relative phase in a larger configuration space.  Even if this procedure is exactly correct, then the only \textit{meaningless} phase is on the universal wavefunction, not individual spin states.  And if one is looking for a spacetime-based reality that might underlie multi-particle configuration space, the empirical accessibility of relative phase implies $\theta(t)$ is a meaningful (if typically hidden) parameter.  For related reasons, the global phase has previously been proposed as a natural ``hidden variable'' in various approaches  \cite{Pearle,KGWE}.  $\theta(t)$ is also the central hidden variable in this approach, as used in Section V.

\section{Nonclassical Framework}

\subsection{The Null Lagrangian Condition}

If one did not know the Born rule, or have any notion of a probabilistic interpretation of $\psi$, the Schr\"odinger equation would be judged a failure: it predicts a time-evolution of $\psi$ that is not empirically observed on a single-particle basis.  The same is true for Maxwell's equations; they completely fail to explain how a photon can travel from (say) a localized atom to (say) a localized CCD pixel.  And yet, these equations are remarkably accurate in the many-particle/high-field limit; they may fail as a fundamental description, but in most cases act as excellent ``guidelines'' for average behavior.

These (and other) fundamental wave equations can be generated from Hamilton's principle $\delta S\!\!=\!\!0$ (where the action is $S=\int {\cal L} d^4x=\int L dt$, integrated over the [space]-time region in question).  Starting from some Lorentz-invariant Lagrangian density ${\cal L}$, then $\delta S\!\!=\!\!0$ generates the corresponding Euler-Lagrange Equations (ELEs) \cite{footnote1}.  Such ELEs are generally wrong when it comes to specific particles, but are often correct on average.  Single-particle phenomena seem to imply that Hamilton's principle is \textit{not} fundamentally correct, but for some reason happens to yield a reasonable guideline for many-particle behavior.  

The most standard explanation of this fact lies in the $\hbar\to 0$ limit of the path integral; only in this limit does one recover Hamilton's principle and the ELEs.   Still, as was discussed in the opening sections, this reading has not led to a realistic understanding of single-particle behavior in our $\hbar\ne 0$ universe.  If all ``paths'' are possible, there is no way to assign each of them a corresponding positive probability.  So, as an alternative to the path integral, one might be tempted to consider theories in which \textit{not} all field configurations were possible, due to some local restriction.  Ideally this restriction would eliminate the ``negative probability'' cases outright.

Looking to (\ref{eq:SaS}) might suggest restricting the path integral to only cases where $S=0$, which would make every term positive and equal.  But this condition is inherently nonlocal, and also strongly depends on what size system one is considering (which should not impact a realistic framework).  Still, there is a related local condition that is as Lorentz-invariant as ${\cal L}$ itself: namely, ${\cal L}\!=\!0$.  This is worthy of preliminary consideration, and will be referred to as the ``Null Lagrangian Condition'' (NLC).  Specifically:

\begin{quote} \textbf{NLC: Apart from external constraints, the only restriction on field variables is that the Lagrangian density is always zero.}
\end{quote}

Under the NLC, $\delta S\!\!=\!\!0$ is abandoned in favor of ${\cal L}\!\!=\!\!0$, and there are no ``law-like'' ELEs.  The NLC is a single scalar equation, so for a Lagrangian density built from at least two scalar fields the allowed behavior is generally underconstrained (although not \textit{so} underconstrained as the usual ``sum over all histories'').  By not imposing $\delta S\!\!=\!\!0$, the NLC permits non-ELE behavior, in concordance with observed single-particle phenomena.  

It is notable that the histories compatible with the NLC are not uniformly spread over parameter space.  For instance, if a solution to the ELEs also obeys ${\cal L}\!\!=\!\!0$, one can typically expect to find a large density of additional  ${\cal L}\!\!=\!\!0$ histories nearby.  This follows from the fact that $\delta' S\!\!=\!\!0$ for any particular variation $\delta'$ that takes one ${\cal L}\!\!=\!\!0$ history to another, and $\delta S\!\!=\!\!0$ holds generally on the ELEs.  One is less likely to find a high density of ${\cal L}\!\!=\!\!0$ histories away from the ELEs, because if $\delta S\!\!\ne\!\!0$ such a variation is not even a candidate for shifting to another ${\cal L}\!\!=\!\!0$ history.  (Although note that even on ${\cal L}\!\!=\!\!0$ ELEs, $\delta S\!\!=\!\!0$ certainly does not imply that \textit{all} neighboring histories obey  ${\cal L}\!\!=\!\!0$; $S$ is an integral, not a local value.)

There are potential objections to the NLC, some of which will be discussed in the final section, but addressing them at this point would be premature.  At this stage, the question is what the NLC accomplishes for the toy Lagrangian of Section III.  

\subsection {A Priori Probabilities}

In the context of the framework (\ref{eq:JPD}), the NLC together with the particular form of ${\cal L}$ defines the allowable microhistories $\mu_j$, but says nothing about the probability of each microhistory $p(\mu)$. Of course, for any given set of measurements/constraints $C$, some microhistories will not be compatible with $C$, and will have a probability of zero.  (Or, alternatively, they are simply not summed over when calculating $P(C)$, which amounts to the same thing if one is careful about normalization.)  The question of what to use for the nonzero $p(\mu)$ is the last major piece of the realistic framework.

One might be tempted to look to $P_{fpi}(C)$ from (\ref{eq:FPI}) in order to guess an appropriate $p(\mu)$, but this is of the wrong form to compare to (\ref{eq:JPD}).  Instead, imposing the NLC on the more comparable form (\ref{eq:SaS}) implies $S=0$, such that all microhistories should have equal $p(\mu)$'s.  Whether or not the assumptions behind (\ref{eq:SaS}) are justified in a field framework, this result is reminiscent of the fundamental postulate of statistical mechanics.  With this in mind one can ignore the FPI entirely, think of $\mu$ as a four-dimensional microstate, and fall back on statistical mechanics and the equal \textit{a priori} probability (EAP) of all microstates.  Extending this to microhistories then simply reads:

\begin{quote}
\textbf{EAP: All allowable microhistories $\mu$ have an equal \textit{a priori} probability.}
\end{quote}

Together, the NLC and the EAP define a new, nonclassical, realistic framework under which one can analyze the Lagrangian from the previous section (and indeed any Lagrangian density). Apart from the details of the Lagrangian, these starting points are as conceptually simple as is reasonably possible.  The only conceivable simpler starting point would be to relax the NLC and consider \textit{all} microhistories, which would put one back in the domain of the (unrealistic) FPI.

Given that the NLC is an underconstraint on the fields that make up ${\cal L}$, it is clear that one cannot typically make exact predictions for any given system.  Indeed, there are no deterministic equations to be solved, there is no Cauchy-problem to pose.  But just as statistical mechanics can yield the deterministic-looking equations of thermodynamics in the many-particle limit, one can use the regions of parameter space with the largest densities of ${\cal L}\!=\!0$ histories to make \textit{a priori} predictions of the average behavior of the fields.  

As noted above, in the special cases when the ELE obeys ${\cal L}\!=\!0$, one can expect to find a large density of ${\cal L}\!=\!0$ histories in the immediate neighborhood of the ELE solutions.  It then seems to be a reasonable (although as yet unproven) conjecture that given no other information, the best guess one can make for the behavior of the fields is that they will track the classical ELEs for which ${\cal L}\!=\!0$.  Even on a single-particle basis, this is a reasonable starting point; there are non-ELE histories with ${\cal L}\!=\!0$, but because $\delta S\ne 0$, deviations around these non-ELE histories are much less likely to encounter other ${\cal L}\!=\!0$ histories.  And because the parameter space away from the ELEs is less dense with NLC histories (as compared to the space near the ELE solutions), the EAP implies that near-ELE histories are far more likely.  (Of course, there maybe \textit{many} unlikely ways that the ELEs might fail, leading to a large net likelihood that the ELEs will fail \textit{somewhere}, but this is precisely the behavior that we are looking for.)

For the toy Lagrangian from section III, where ${\cal L}\to L$, a qualitative schematic of the $L=0$ microhistories is shown in Figure 1.  The thick lines represent the ELEs, while the thin lines are other $L=0$ histories.  All things being equal, there is a higher density of $L=0$ histories near an ELE solution, so if it is known that the starting point is $A$, ones best guess for an outcome would be $Y$.  But because the ELEs are not ``laws'', $Y$ need not be the actual outcome.  If it is not, the EAP can be used to determine which outcome is more likely, and indeed calculate the relative probabilities (as will be done quantitatively in section V).  In Figure 1, given the partial constraint A, it should be clear that the outcome $X$ is more likely than $Z$.

\begin{figure}[htb]
\centerline{\includegraphics[width=.5\textwidth]{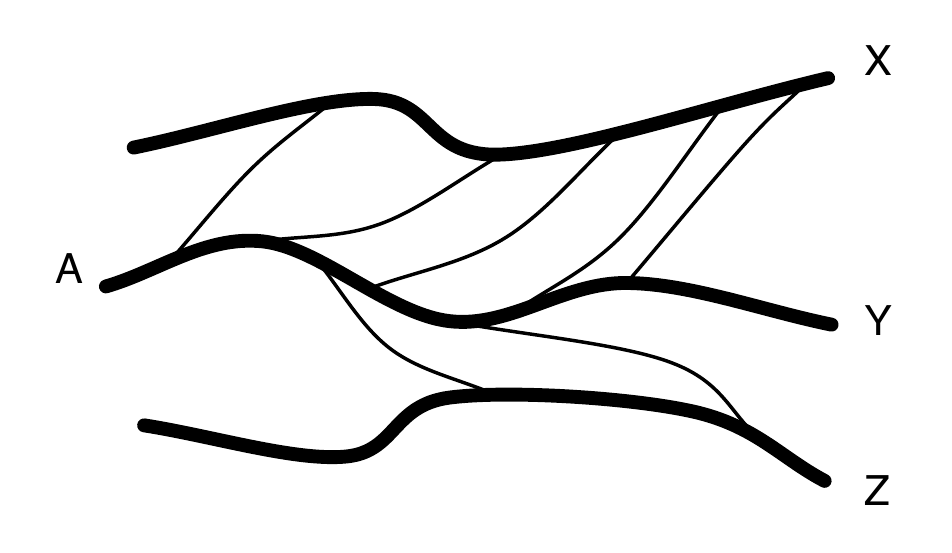}}
\caption{A qualitative schematic of the space of $L=0$ microhistories; time runs horizontally.  The thick lines are solutions to the Euler-Lagrange equations (that also obey $L=0$), and closely neighboring $L=0$ histories.  The thin lines are other $L=0$ histories.  Given an initial state of $A$, and no other information, the EAP implies that the final state will probably be $Y$.  Given that the final state is not $Y$, the EAP implies that $X$ is more likely than $Z$.}
\label{Figure:fig1}
\end{figure}

\subsection {High-Probability Measurements}

Section III demonstrated that the classical solution for the toy Lagrangian exactly maps to that of an unmeasured quantum spin state.  But it does not map to that of a \textit{measured} spin state; there is no collapse mechanism to select energy eigenvalues as discrete outcomes.  Remarkably, the framework from this section can effectively select for these outcomes with only one additional concept: the time-energy uncertainty principle.

In CQM, the time-energy uncertainty principle looks quite different from the position-momentum uncertainty principle, primarily because there is no time operator.  And yet, relativity demands a clear link between the two, as does basic Fourier analysis (assuming the standard connection between frequency and energy).  Still, the empirical fact of the time-energy uncertainty principle is sometimes ignored in CQM by theoretically imposing an arbitrarily-accurate energy measurement at an arbitrarily-accurate moment.  But even though one can do this on paper, one cannot do it in practice: if precise energy measurements occur at some particular time, that time cannot be controlled or known to better than $\hbar/\Delta E$.  Whether or not CQM can adequately explain this, it must be considered when modeling energy measurements.

With this empirical uncertainty in measurement time, microhistories which are highly periodic over the uncertainty window become significantly more probable than non-periodic microhistories.  This follows from the EAP almost regardless of how the measurement is actually imposed on $\ket{q(t)}$.  In the simplest case, the energy measurement at $t\!\!=\!\!t_f$ can be modeled as a hard absorption of $\ket{q(t)}$ (into some external device) such that $\ket{q(t\ge t_f)}=0$.  For continuity of $\ket{q}$, the phase would be fixed at $t_f$, like the temporal equivalent of the familiar spatial scenario where an electromagnetic field terminates at a perfect conductor.  By virtue of the interaction, the measurement device would then become correlated to values of both $q(t_f)$ and $\dot{q}(t_f)$, from which the device could compute the frequency/energy of the system at the time of measurement.

The key point is that for a given set of measured values $q(t_f)$ and $\dot{q}(t_f)$, the value of $t_f$ itself is unknown, due to the time-energy uncertainty principle.  And if there are N different microhistories that might yield those \textit{same} measured values, terminating at N different values of $t_f$, then the EAP implies the corresponding outcomes are N times more probable than single-microhistory outcomes.  This strongly selects for periodic solutions.  For non-periodic solutions, any given values of $q(t_f)$ and $\dot{q}(t_f)$ are liable to only occur at a very few values of $t_f$, and will be correspondingly less likely.  

One can even estimate the value of $N$.  The natural periodicity of the system from section III is on the order of the Compton period, $h/(mc^2)$.  The longest periods occur for the lightest particles; for an electron this is $\approx10^{-20}$s.  If one localizes the timing of the energy measurement to within 1 ns, $N=10^{11}$.

It might seem that this solution to recovering discrete outcomes would not extend to other types of measurements, but to address that issue one would need to reintroduce the spatial components of the Lagrangian density.  At that point, other periodicities come into play; if a momentum measurement occurs at some unknown location, the above analysis would strongly select periodic solutions \textit{in space}, exactly the sort of momentum eigenstates one would expect.  Other measurements seem amenable to this treatment as well \cite{Wharton09}, but such issues are beyond the scope of this paper.

The net result of this analysis is that the fields that make up the toy Lagrangian are overwhelmingly likely to be found in high-frequency exactly-periodic modes at the time of measurement, which happen to correspond to energy eigenstates.  In a fully classical system, this result would be irrelevant, as the mode at the time of measurement would not be free; it would be fully determined by the initial state of the system and the ELEs.  But in this case, there are no strict laws that govern the dynamics; there are $L=0$ paths connecting different ELE solutions, as shown schematically in Figure 1.  In that diagram, if X and Z were highly-periodic eigenstates, and Y was not periodic, then (given a sufficient number of $L=0$ paths between ELEs) X and Z would become the only observable outcomes, for all practical purposes.  This qualitatively matches empirical observations; it now remains to be seen whether the correct probabilities for energy measurements of spin states in arbitrary magnetic fields can be quantitatively deduced from this framework.

\section{Applications}

\subsection{Deriving the Born Rule}

The central result of this paper is the following application of the section IV framework to the toy Lagrangian of section III.  This allows one to derive the expected outcome probabilities from the relative numbers of NLC histories, subject to any particular set of external constraints.  Several assumptions will be made when first deriving the probabilities, but these assumptions will be revisited immediately afterward.

Given that a measurement at $t\!\!=\!\!t_f$ is overwhelmingly likely to result in an energy eigenstate, it is convenient to parameterize $\ket{q}$ in the basis defined by the eigenstates of $-\gamma\bm{S\cdot B}(t_f)$.  In this basis, for times near $t_f$ (in the limit that the magnetic field is constant), we can always write the $n$ complex components of $\ket{q}$ as:
\begin{align}
\label{eq:param}
q_1&=A (cos\,\alpha) e^{-iE_1t/\hbar}e^{i\theta}, \nonumber \\
q_{j\ne 1}&=A (sin\,\alpha) c_j e^{-iE_jt/\hbar}e^{i\theta}.
\end{align}
Here the $n\!-\!1$ complex parameters $c_j$ as well as the three real parameters $A$, $\alpha$ and $\theta$ are all functions of time (there is no parameter $c_1$, as it can be absorbed into $\alpha$ and $\theta$).  The values $E_j$ are the corresponding eigenvalues of $-\gamma\bm{S\cdot B}(t_f)$.

This is still too many parameters to easily consider all possible $L=0$ histories, but one can make progress by forcing $\dot{c}_j=0$, as would be the case for solutions to the ELEs.  Some non-ELE behavior of the $q_{j\ne1}$ terms is still permitted, via $\alpha$, the global phase $\theta$, and the global normalization $A$, but this approximation will clearly hide any nonclassical excursions within the the $c_j$'s themselves.  (We shall see that it is possible to come to solid conclusions without such details.)  Under this assumption, the $n\!-\!1$ values of $c_j$ can be independently normalized such that the sum of $|c_{j}|^2$ equals 1.  Therefore, if $\ip{q}{q}=1$, it follows that $A=1$ as well; we can impose this as a boundary constraint at every measurement (but not between measurements).

Inserting this form of $\ket{q}$ into the NLC equation $L=0$, yields
\begin{equation}
\label{eq:NLC}
m^2c^4-(\hbar \dot{\theta})^2  - \frac{(\hbar \dot{A})^2}{A^2} - (\hbar \dot{\alpha})^2 = 0.
\end{equation}
Since the ELEs obey $L=0$, the classical time-evolution of $\ket{q}$ will solve (\ref{eq:NLC}); in this classical limit, $\dot{\theta}=\dot{\theta}_0\equiv -mc^2/\hbar$ and $\dot{\alpha}=\dot{A}=0$.  But if $\ket{q}$ is not an energy eigenstate, it will not be highly periodic, and the analysis from the previous section implies that the ELEs will not yield the most probable history.  To shift into an eigenstate, it is necessary to briefly depart from the ELEs, allowing $\dot{\alpha}$ to be nonzero for some portion of time.  During that portion, the NLC implies that (\ref{eq:NLC}) must still hold.

Given (\ref{eq:NLC}), the only way for $\dot{\alpha}$ to become nonzero is for $|\dot{\theta}|$ to also become lower than the classical value $|\dot{\theta}_0|$.  Notice that nonzero values of $\dot{A}$ cannot substitute for $\dot{\theta}$; if $\dot{A}$ varies, $\dot{\theta}$ will have to further adjust to maintain the NLC.  Seeing that nonzero values of $\dot{A}$ are neither necessary nor sufficient to shift to an eigenstate, and that excursions away from the ELE solutions carry a probabilistic penalty via the EAP, it is reasonable to maintain $A=1$ at all times.  (This may be as close as one can come to an analog for energy conservation in this framework.)

Defining the anomalous global phase change $\dot{\theta}_a\equiv\dot{\theta}-\dot{\theta}_0$, where $|\dot{\theta}_a|\ll|\dot{\theta}_0|$, then reduces (\ref{eq:NLC}) to a simple relationship between $\dot{\alpha}$ and $\dot{\theta}_a$.  Time averaging this result over the relevant duration $t_0$ (the timespan over which the non-ELE anomalies might occur) yields
\begin{equation}
\label{eq:NLC2}
<\!\dot{\theta}_a\!> = \frac{<\!\dot{\alpha}^2\!>}{|2\dot{\theta}_0|}= \frac{\sigma^2+<\!\dot{\alpha}\!>^2}{|2\dot{\theta}_0|}.
\end{equation}
Here $\sigma$ is the standard deviation of $\dot{\alpha}$ (away from zero, its classical value).  As there is only one remaining unconstrained parameter in $\ket{q}$, $\sigma$ is a quantitative metric for how ``nonclassical'' the history $\ket{q}$ is, and the product $\sigma t_0$ is therefore a measure of the net deviation from the ELEs.  (Leaving one ELE solution and joining another requires $\dot{\alpha}$ to become nonzero for a time, before returning to zero.)  It may be that a careful analysis of the $L=0$ density could be used with the EAP to actually \textit{solve} for the \textit{a priori} probability $P(\sigma t_0)$, but that calculation has not yet been performed.  For this proof-of-principle, it seems reasonable to guess that the density of $L=0$ histories will be normally distributed around $\sigma t_0=0$ (the ELEs), with the characteristic scale to the gaussian defined here as the unknown parameter $\gamma$.

Rewriting (\ref{eq:NLC}) in terms of the net global phase anomaly $\theta_a=t_0<\!\dot{\theta}_a\!> $ and the net anomalous rotation of the spin state $\alpha_a = t_0<\!\dot{\alpha}\!>$, and then convolving with the normal distribution $exp(-\sigma^2t^2_0/2\gamma^2)$, yields 
\begin{equation}
\label{eq:NLC3}
\theta_a(\alpha_a) = \frac{\gamma^2+\alpha^2_a}{|2\dot{\theta}_0|t_0}.
\end{equation}
Notice that the global phase is extremely hard to shift beyond its classical value (as determined by the ELEs); even with a large nonclassical shift in the angle $\alpha$, the product $|\dot{\theta}_0|t_0$ is at least as large as N, if not much larger.  Also, note that the inherent $\Delta t$ uncertainty in $t_0$ (due to the time-energy uncertainty principle) implies an inherent uncertainty in the global phase.  (For $\gamma^2\ll\alpha_a^2$, this uncertainty dominates a small additional uncertainty due to the convolution.)  If $\Delta t \ll t_0$, the uncertainty in the global phase is
\begin{equation}
\label{eq:NLC4}
\Delta \theta (\alpha_a) = \frac{\Delta t}{|2\dot{\theta}_0|t_0^2} \left( \gamma^2 + \alpha^2_a\right).
\end{equation}

The final issue is how to utilize the master probability equation (\ref{eq:JPD}), which means determining what external constraints $C$ are imposed on the global phase $\theta$ when an energy measurement is made.  At first glance the global phase might appear to be irrelevant -- after all, the uncertainty in the measurement time $\Delta t$ is much larger than $1/|\dot{\theta}_0|$.  But one should not fall into the trap of thinking of the measurement device as a purely classical system; if this framework truly underlies quantum systems, it must underly macroscopic systems as well.  This implies that the measurement device has its \textit{own} global phase $\theta'$, naturally oscillating at a much larger frequency $\dot{\theta}_0'\equiv -Mc^2/\hbar$ (where M is the mass of the measurement device).  Furthermore, however improbable it is to anomalously change the phase of the spin-state, it is even more improbable to change the phase of $\theta'$; a comparable analysis to (\ref{eq:NLC3}) for the measurement device would yield $\dot{\theta}_0'$ in the denominator.

Therefore, if there is any phase-matching between the measuring and measured systems (as there always is when oscillators couple to each other), the large value of $M/m$ will effectively act as an external constraint on the global phase $\theta$ of the spin state.  (There is an interesting exception to this, discussed in the next subsection.)

This analysis implies that the spread of global phases $\Delta \theta$ over different microhistories $\ket{q(t)}$ at the time of measurement dilutes the probability of any particular externally-constrained phase.  (In the limit that $M/m\to\infty$, the external constraint $C$ imposed by the measurement device will only be compatible with one particular value of $\theta(t_f)$.)  So when summing over only the allowable microhistories, the relative probability of a correct phase-match will scale like $1/\Delta \theta$, or
\begin{equation}
\label{eq:Born}
P_0(\alpha_a) \propto \frac {1}{ \gamma^2 + \alpha^2_a}.
\end{equation}
In the limit that $\gamma\to0$, this is precisely the distribution of non-classical behavior needed to reproduce the Born rule, as shown by Schulman \cite{Schulman}.  (Although note that one difference here is that $\alpha_a$ is the integrated anomaly, not a series of independent ``kicks''.)

Following Schulman, the final step in the derivation runs as follows.  Given that $\ket{q}$ is overwhelmingly likely to end up in an energy eigenstate, $|q_1|$ must become either 1 or 0.  Since $A$ is fixed at 1, this requires an anomalous change of the angle $\alpha$ of either $\alpha_a = l\pi-\alpha$ (for $|q_1|=1$) or $\alpha_a=l\pi+\pi/2-\alpha$ (for $q_1=0$), where $l$ is an integer.  Because one of these outcomes must occur, the ratio of the probability of these two cases is
\begin{equation}
\label{eq:cot2}
\frac{P(|q_1|=1)}{1-P(|q_1|=1)}=\frac{ \displaystyle\sum_{l=-\infty}^{\infty}P_0(l\pi-\alpha)}{\displaystyle\sum_{l=-\infty}^{\infty}P_0(l\pi+\pi/2-\alpha)}=cot^2\alpha.
\end{equation}
The final result of $cot^2(\alpha)$ comes from inserting (\ref{eq:Born}) as the form of $P_0$, and taking the limit as $\gamma\to0$.  From (\ref{eq:cot2}), it is obvious that $P(|q_1|=1)=cos^2(\alpha)$, exactly the same probability as one would get from using the Born rule on (\ref{eq:param}) with no anomaly at all.  Because $q_1$ could have corresponded to \textit{any} eigenstate, this result holds for \textit{every} eigenstate, and the Born rule is recovered -- at least to the limits of the assumptions used above.  As the next subsection demonstrates, these limits can be expanded significantly, but fail in interesting cases.   

\subsection{Extensions and Exceptions}

One of the most important reasons to pursue theories that might underly CQM is as a path to new physical predictions.  For example, without Boltzmann's realization that the statistics of microstates underlies (macroscopic) entropy, it would have been impossible to predict that the second law of thermodynamics is expected to occasionally fail (on short timescales for small systems).  In the same way, any theory that underlies CQM should reveal regimes in which CQM would be expected to fail.  With this in mind, the various assumptions in the preceding derivation demand serious attention.

In deriving the Born rule for spin states, the first set of assumptions was that the magnetic field was the constant $\bm{B}(t_f)$, its value at the time of measurement.  But if anomalous (non-ELE) behavior is possible at times near $t_f$, it is also possible at earlier times.  At these earlier times, if the magnetic field is allowed to vary, there is no justification for setting $\dot{c}_j=0$, or indeed writing $\ket{q(t)}$ in the form (\ref{eq:param}).  The relevant basis for a more general anomaly at time $t\ne t_f$ is not the energy eigenstates at that earlier time.  Instead, it is determined by the entire magnetic field history $\bm{B}(t)$ to $\bm{B}(t_f)$: taking the final eigenstates at $t_f$, one can then use the ELEs to backward-time-evolve those states to an arbitrary time $t$.  This is the relevant basis $q_j(t)$, because if an anomaly allowed $\ket{q(t)}$ to transition into one of these states (say, $q_1(t)=1$, $q_{j\ne1}(t)=0$), then the usual ELE dynamics will naturally bring this non-eigenstate into the high-probability (periodic) eigenstate at the future measurement time $t_f$.  Schulman calls these $q_j(t)$ ``special states'', as defined by their special future behavior \cite{Schulman}.

In the basis of these ``special states'', one cannot generally use the form (\ref{eq:param}).  Instead, the more-general form
\begin{align}
\label{eq:param2}
q_1&=A (cos\,\alpha) e^{i\theta}, \nonumber \\
q_{j\ne 1}&=A (sin\,\alpha) c_j e^{i\theta},
\end{align}
restores $c_j$ to a function of time because  $\dot{c}_j=0$ does not conform to the ELEs.  Despite this difference, one can still set $c_j(t)$ to the non-anomalous value it would have under the ELEs, again restricting the entire non-ELE anomaly to the three parameters $\theta$, $\alpha$, and $A$.  In this case, (\ref{eq:NLC}) is no longer exact; the calculation of $L=0$ introduces new terms, the largest of which is $<\!q|(\gamma \bm{S\cdot B})|q>\hbar\dot{\theta_a}$.  But in the non-relativistic energy limit that all of the eigenvalues $|E_j|\ll mc^2$, this is always negligible compared to the dominant term $mc^2-(\hbar \dot{\theta})^2\approx 2mc^2\hbar\dot{\theta_a}$, and the above analysis still holds good.

This generalization implies that the anomalous behavior of $\ket{q(t)}$ can occur at any time between the preparation and the measurement, not limited to a ``collapse''-like behavior right at $t_f$.  The duration $t_0$ of possible anomalies therefore can be associated with the time between measurements (and $\Delta t$ is the uncertainty in $t_0$).  The probability distribution over the outcome is still governed by the Born rule, at least in the non-relativistic limit.

Note that this extension still utilizes the assumption that all of the non-ELE anomaly can be encoded in $\alpha$ and $\theta$, and ignores any potentially anomalous behavior of the $c_j$'s.  But because of the way that the probabilities are calculated in (\ref{eq:cot2}), any non-ELE anomaly in the sector orthogonal to $q_1$ is irrelevant to that particular ratio of probabilities; $|q_1|$ must still go to 1 or 0, and the $c_j$'s cannot contribute either way.

Going through the other assumptions, the most solid one is $|\dot{\theta}_a|\ll|\dot{\theta}_0|$.  Because of (\ref{eq:NLC}), if $|\dot{\theta}_a|$ approaches $|\dot{\theta}_0|$, then it also becomes comparable to $\dot{\alpha}$.  But during such a large anomaly, the net anomalous angles must also be comparable, $\theta_a \approx \alpha_a$.  This anomalous change in the global phase is at least a factor of $N$ bigger than Eqn. (\ref{eq:NLC3}) for the same $\alpha_a$, and since $P_0 \propto 1/\Delta\theta$, this enormous (relative) change in $\theta$ is very unlikely.

A more interesting assumption is that $\gamma$ is small, but nonzero.  Schulman has calculated the expected deviations from the Born rule due to a finite $\gamma$, but stressed that this was not a testable prediction because he had no way to estimate its value \cite{Schulman}.  Here, however, $\gamma$ is defined in terms of the density of $L=0$ histories, and could in principle be calculated.  If it was found to be sufficiently large, then (\ref{eq:cot2}) could noticeably deviate from $cot^2(\alpha)$, and experiments could contrast the probabilistic predictions of this approach with CQM's Born rule.  On the other hand, if $\gamma$ was found to be sufficiently close to zero, then the penalty for non-ELE excursions might be so high that in some cases this approach might predict a non-negligible probability of measuring an outcome that was \textit{not} an energy eigenstate.  (The key parameter in that case would be $N=|\dot{\theta}_0|\Delta t $, where smaller $N$'s would make non-eigenstate-outcomes more likely.) Such outcomes are usually discarded as experimental error (as there is no framework that predicts their occurrence), but if there was a prediction for their likelihood, that could also be tested.

Another assumption is that $\Delta t \ll t_0$, meaning the uncertainty in the measurement time is much less than the time between measurements.  But even if this limit is not valid, $\Delta\theta_a(\alpha_a)$ is still proportional to $\gamma^2+\alpha^2_a$, and one recovers (\ref{eq:Born}).  Nevertheless, it is possible that experimental tests can be devised if different outcomes can somehow be associated with different $\Delta t$'s; in this case (\ref{eq:Born}) would no longer follow from (\ref{eq:NLC4}). 

The final assumption in the derivation was that the phase of the measurement device $\theta'$ was constrained according to roughly the same procedure as the system it is measuring.  Namely, that the measurement device ($M$) would \textit{itself} be eventually measured by a larger-mass detector ($M'$), fixing its phase by an even stricter constraint.  But in principle, this depends on the experiment; if the phase information in $M$ is carefully manipulated, it can be effectively erased, voiding the constraint on $\theta'$ and therefore changing the expected outcomes on the original system. 

Remarkably, this scenario is perfectly compatible with known results of quantum eraser experiments \cite{eraser}.  If the details of the measurement never become accessible to the rest of the universe (and to any cosmological boundary conditions), then there is no ultimate constraint on the quantum system, and the section IV logic that demanded an eigenstate-outcome falls apart  \cite{FQXi3}.  With no constraint on the outcomes, one would revert to the \textit{a priori} guess of the ELEs, as is known to be the correct procedure when a measurement is erased.  This realistic account of quantum eraser experiments is not available to the instantaneous-states of CQM, because at the time of measurement there is no way of knowing whether the correlations imposed by the measurement are in turn correlated with anything else.  But in this 4D-microhistory perspective, one takes a longer view, and future events can indeed constrain the expected behavior at earlier times.

\subsection{Realistic Entanglement}

Looking to the future to explain the past behavior of delayed-choice experiments is an approach that has also been noted as a potential realistic account of quantum entanglement, Bell's theorem notwithstanding \cite{OCB,CWR,Cramer,Sutherland,Price,Miller,Wharton10,Aharonov}.  As discussed in section II, such no-go theorems always rule out future constraints on past behavior by fiat, despite the fact that the FPI mathematics imposes exactly such a constraint.  The nonclassical framework from section IV is different from FPI, but effectively gets around such theorems in the same manner.

What has been missing in these prior arguments, however, has been an explicit realistic model of entanglement -- one which can answer quantitative questions about the system between measurements, as a function of space and time \cite{footnote2}.  With the above results, such a model is now possible.  Although there is no spatial extent to the toy Lagrangian of section III, the implication was that the classical field variables $\ket{q(t)}$ are associated with the spin state of a particular particle, and can therefore be assigned to the that particle's location (to the extent that it is known).  A two-particle system ($L$ and $R$) can have their spin-states represented by $\ket{q_L(t)}$ and $\ket{q_R(t)}$; even though there is no explicit spatial dependance, the variables $\ket{q_L(t)}$ can be associated with the general spatial location of particle $L$ at time $t$ (and the same for $R$).  This spin-state model cannot be applied to which-way entanglement, but can be used for scenarios where there is no doubt about the general trajectory of each particle.

Extending the 1-particle model to a 2-particle model is most naturally accomplished using the ``action duality'' technique developed in \cite{WMP} and \cite{EPW}.  First, consider a specific 1-particle experiment.  In figure 2a, a single spin-1/2 particle undergoes an energy measuement ($M_1$) at time $t=-t_f$.  During the measurement, there is a magnetic field aligned at a $2\alpha_0$ angle from the z-axis (towards the x-axis).  By selecting for one particular outcome, and then adiabatically turning off the B-field, this can result in a known state at $M_1$ of
\begin{equation}
\label{eq:q0}
\ket{q_1(t)}=\left( \!\! \begin{array}{c} {cos\, \alpha_0} \\ {sin \,\alpha_0} \end{array} \!\! \right)e^{i(\theta_i-mc^2t/\hbar) }.
\end{equation}
Here $\theta_i$ is an unknown (but still constrained) global phase.  This is the \textit{a priori} best guess of what will happen in the future of $M_1$ (the ELE solution).  

Later, at time $t=+t_f$, a z-directed magnetic field is turned on and an energy measurement is made ($M_2$), resulting in one of two outcomes.  From this one can make an \textit{a posteriori} guess of what has happened in the past of $M_2$ (the ELE solution, given only the outcome).  For the spin-up outcome, one would guess
\begin{equation}
\label{eq:qplus}
\ket{q_{2+}(t)}=\left( \!\! \begin{array}{c} {1} \\ {0} \end{array} \!\! \right)e^{i(\theta_f-mc^2t/\hbar) };
\end{equation}
for other outcome, $\ket{q_{2-}(t)}$, would use $(0 \,1)$ instead of $(1 \, 0)$.  This solution is again constrained by an unknown global phase $\theta_f$.  Note that while one has effective experimental control over the preparation at $t\!=\!-t_f$ (via preselection), there is no control over the outcome of the final measurement at $t\!=\!+t_f$ (except via postselection, which is usually not considered to be ``control''). 

The previous analysis yields a probabilistic description of what actually happens to $\ket{q}$ between the two measurements.   Even with a spin-up outcome, both (\ref{eq:q0}) and (\ref{eq:qplus}) are incorrect guesses; the actual history will be constrained by both $M_1$ and $M_2$, and has the form
\begin{equation}
\label{eq:qa}
\ket{q(t)}=\left( \!\! \begin{array}{c} {cos\, \alpha(t)} \\ {sin \,\alpha(t)} \end{array} \!\! \right)e^{i\theta(t) }.
\end{equation}
The NLC correlates $\alpha(t)$ with $\theta(t)$, and both functions are constrained at two times.  Specifically, $\ket{q(-t_f)}=\ket{q_1(-t_f)}$ and $\ket{q(t_f)}=\ket{q_{2\pm}(t_f)}$ (where the sign is linked to the outcome).  Applying the EAP to the two classes of possible histories (\ref{eq:qa}) then yields the ratio of probabilities $P(q_+)/P(q_-)=cot^2\alpha_0$, according to (\ref{eq:cot2}).  Once the outcome is known, all consistent histories $\ket{q}$ are equally probable (in hindsight), and exactly one of these can be said to be the actual history.  

\begin{figure}[htb]
\centerline{\includegraphics[width=.5\textwidth]{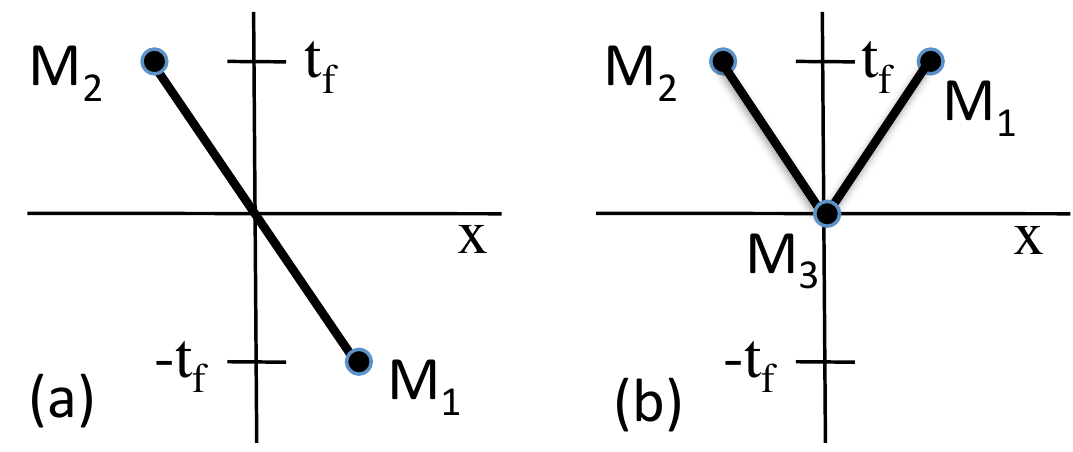}}
\caption{Two action-dual experiments.  (a) A single spin-state is prepared at M1 and measured at M2.  (b) Two spin-states are jointly prepared/correlated at M3, then independently measured at M1 and M2.  Note that (a) and (b) are related by time-reversing the events on the right-hand side.}
\label{Figure:fig2}
\end{figure}

One advantage to realistic spacetime descriptions is that they can be used to analyze dual experiments (with different geometries) using simple transformations.  Specifically, consider Figure 2b; here the first half of the original trajectory is time-reversed around $t=0$, creating a 2-particle geometry.  $M_1$ is now an outcome measurement (rather than a preparation) and one loses apparent control over the outcome.  Assuming for now that the same outcome is postselected, $M_1$ constraints an almost-identical state as $\ket{q_1}$ from (\ref{eq:q0}); the only difference is that $t\to-t$ as per the time-reversal.  This means that the new best-guess (given only $M_1$) for the particle on the right is $\ket{q_R(t)}=\ket{q_1(-t)}$; this still solves the classical ELEs, but as the previously-neglected antimatter-like solution ($\ket{q_-}$ from section III.B).  The best guess for the particle on the left, given only $M_2$, is still $\ket{q_{L\pm}(t)}=\ket{q_{2\pm}(t)}$.

The only other change is that there appears to be a 3rd measurement, $M_3$.  This is a joint measurement on the (now) 2-particle system that simply imposes the same constraint that mere continuity imposed in figure 2a at the same spacetime location.  Namely, it is the constraints $\ket{q_L(0)}=\ket{q_R(0)}$ and $\ket{\dot{q}_L(0)}=-\ket{\dot{q}_R(0)}$.  From this constraint at $M_3$, one can even construct a best-guess of $\ket{q_{L}(t)}$ given only $M_1$; it would be identical to $\ket{q_1(t)}$.  One can also make a guess for the state of both particles given only the outcome at $M_2$.  As before, these guesses simply use the ELEs to extrapolate from what is given, but all of these guesses are incorrect (except in the very special case where an ELE solution actually matches both boundary conditions at $M_1$ and $M_2$).

By building an entangled model in this manner, the actual field variables at all times can be constructed from (\ref{eq:qa}), via the same transformation that takes Figure 2a to 2b.  Applying the EAP then trivially yields the same (joint) probabilities for the 1-particle measurements $(M_1,M_2)$ and the 2-particle measurements $(M_1,M_2,M_3)$, so long as $M_3$ is assumed to agree with the continuity of the 1-particle case.  If this $M_3$ is taken as a given preparation, and $M_1$ is postselected to match the preparation of the 1-particle case, then the probabilities of the outcome at $M_2$ are again $P(q_{L+})/P(q_{L-})=cot^2\alpha_0$.  

But in the 2-particle case, there is no reason to postselect on one of the outcomes at $M_1$.  Given the joint preparation $M_3$, the EAP can yield the joint probabilities for all possible outcomes $(M_1,M_2)$, and they can be computed for any value of $2\alpha_0$ (the orientation of the magnetic field at $M_1$).  Typically we take measurement settings like the value of $\alpha_0$ to be constraints under our control (call such constraints $C'$, a subset of the external constraints $C$), and we normalize (\ref{eq:JPD}) such that $P(C')=1$.  (Meaning we don't even consider histories that violate $C'$.)  We typically do not take the outcomes to be under our control, even though these are also part of $C$; until we learn what they are, we take their probabilities to be distributed according to $P(C)$. (Such subjective probabilities are then updated upon gaining new information, as is usual.)

Where the Bell-inequality violations occur is when we compare different experiments, with different magnetic field orientations $\alpha_0$.  But under such a comparison, one must consider entirely different microhistories, as is clearly evident in this framework.  To change $\alpha_0$ one first alters the controllable constraints $C'$ on the microhistories, renormalizes all the probabilities such that $P(C')$ is again unity, and only then recomputes the outcome probabilities.  In this process, certain classes of histories (Schulman's ``special states") become N times more probable than they were in the original experiment (because there are now N times as many corresponding microhistories resulting in indistinguishable outcomes).  The fact that histories span the entire duration of the experiment makes it explicit that one can have realistic accounts of different experiments that need not map to each other at \textit{any} time -- even if the only difference between them is a setting $\alpha_0$ that exists in the future.    

Allowing such a counterfactual ``variation" of $\alpha_0$, the joint probability distribution $P(M_1,M_2)$ violates the Bell inequality, even though it is explicitly generated from a realistic hidden variable model.  Again, this is permitted because the actual history (\ref{eq:qa}) is constrained by future events (the settings at $M_1$ and $M_2$, which select the high-probability outcomes via the EAP), and such future constraints on past behavior are (implicitly or explicitly) assumed to be impossible in any valid derivation of Bell's inequality.

The other notable feature of this 2-particle model is that its underlying parameters are based in spacetime rather than the (much larger) multiparticle configuration space of CQM.  (This was defined as a requirement for a ``realistic theory".)  It is perfectly natural for states of knowledge to exist in a space larger than that of the underlying framework \cite{Spekkens,EPW}, but this fact alone does not prove that spacetime is large enough to hold a realistic basis for CQM's multiparticle configuration space.  While previous arguments have indicated spacetime is likely sufficient \cite{WMP}, this result is an explicit demonstration that such a reduction can actually be accomplished.

\section{Summary and Discussion}

The central novelty of this paper lies in the proposed framework from section IV.  Based upon the success of this framework as applied to the toy Lagrangian of section III, it seems reasonable to elevate the framework to a pair of independent \textit{postulates}, to be evaluated on their own merits and by using better-established Lagrangian densities.
	
The first postulate is the Null Lagrangian Condition (NLC), which simply demands that ${\cal L}\!=\!0$.  While little justification was originally given, this postulate straddles a middle ground between the classical $\delta S\!=\!0$ (Hamilton's principle) and the Feynman Path Integral (FPI) with its quantum field theory extension.  The classical case tells us we should only consider histories that are solutions to the Euler-Lagrange equations (ELEs); the FPI tells us that we should consider all possible histories (ELEs or not).  The NLC allows more histories than $\delta S=0$ but fewer than the FPI.  It is also crucial that the NLC histories are not uniformly distributed over parameter space; if the ELEs solve ${\cal L}\!=\!0$, the NLC histories tend to be clustered around the classical ELEs.  

The second postulate is the Equal \textit{a priori} Probability (EAP) of all microhistories.  This is taken directly from statistical mechanics, shifting the application from 3D states to the 4D field histories that might possibly fill spacetime.  Once we apply our knowledge of any given system as constraints on these histories, ordinary probability theory yields the corresponding predictions via (\ref{eq:JPD}).

At this juncture, the primary support for these postulates lies in the results from section V -- most notably the derivation of the Born rule.  This is not a mere \textit{reinterpretation} of quantum theory, but rather an outline of a proposed \textit{explanation}, in the same way that statistical mechanics is an explanation of thermodynamics.  Such an explanation appears to be testable (at least in principle), as noted by the various exceptions to the Born rule derivation in section V.B.  However, without expanding the toy Lagrangian to a more general Lagrangian density, it remains to be seen whether these postulates are even consistent with known experiments, let alone additional tests.

Before this future stage of research begins, it is appropriate to consider the NLC and EAP on their own merits.  Concerning the technical objections that might be directly raised against the NLC, some may worry that by replacing (\ref{eq:FPI}) there is no clear way to recover classical physics in the $\hbar\to0$ limit.  But whether or not such an unphysical limit is even relevant, the point of the NLC is to deduce \textit{quantum} behavior, and then approximate classical physics from there.  The only reason this two-step approach might be faulted is if there was some deep disconnect between quantum physics and classical physics, and indeed such a disconnect is usually lumped into the so-called ``measurement problem''.  But that is not a critique of the NLC; that is a failure of CQM.  (Also, to the extent that $\hbar$ is even needed \cite{nohbar}, it can appear in ${\cal L}$.) 

Another issue is that ${\cal L}=0$ appears to be violated on macroscopic scales.  For example, the classical Lagrangian for electromagnetic fields is ${\cal L}_{EM}=F^{\mu\nu}F_{\mu\nu}\propto \bm{E}^2-c^2\bm{B}^2$.  This is in fact zero for the classical fields produced by the arbitrary acceleration of a single charged particle, but is not necessarily zero for the superposition of two such fields.  However, note that the NLC does not demand that ${\cal L}_{EM}=0$; rather, it is a constraint on the total Lagrangian density, including interaction terms.  There is interesting research to be done concerning whether ${\cal L}_{EM}\ne0$ can still be compatible with the NLC, by adding in the Einstein-Hilbert Lagrangian density and/or virtual matter fields. Indeed, imposing non-classical constraints on ${\cal L}_{EM}$ may be \textit{required} in order to explain well-known events such as localized photon absorption.  Just as the fields produced by a (single, classical) accelerated charge obey ${\cal L}_{EM}=0$, so do all possible fields that can be completely \textit{absorbed} by a single charge.  This may therefore be an important feature for any eventual Lagrangian-only measurement theory.  

Concerning the EAP, it is unclear whether any technical objections even exist -- and yet, this second postulate is perhaps more likely to be rejected than the NLC.  The issue here (with some physicists) tends to be a philosophical rejection of the block universe framework under which one can even discuss entire microhistories (let alone perform statistical analyses on them).  This rejection persists despite the importance of the block universe in general relativity and classical field theory in general.  But as the block universe is aligned with these well-established physical theories, such concerns are not specific to the EAP.

Another conceptual barrier to the EAP might be the common assumption that the universe fundamentally operates according the ``Newtonian Schema'' of initial states, dynamics, and outcomes \cite{FQXi4}.  Without dynamical laws, the less-intuitive ``Lagrangian Schema'' of the EAP cannot be reduced to this framework.  Perhaps the only reason that general relativity didn't immediately lead to a 4D generalization of 3D statistical mechanics was the assumption that the 3D states were all linked via some deterministic set of equations (making 4D statistics trivially reducible to 3D statistics).  Even today, the Wheeler-DeWitt equation might be used as an argument for only considering probabilities in 3D (where all the freedom lies), rather than in 4D (where nothing interesting happens).  And yet, if there is anything to Feynman's insight of considering histories that do \textit{not} solve any particular dynamical equations, these arguments may fail.  Without well-defined dynamics, one might even argue the 4D EAP is the only natural position to take.

Regardless, it is not unfair to propose that if the only barrier to a line of physical inquiry is a philosophical bias closely linked with our experience of time and causal order \cite{Price}, than one should perhaps attempt to set that bias aside and see what might result.  For the toy Lagrangian of section III, what results is a realistic, continuous, hidden-variable account of what may be happening between spin measurements, even including cases that violate Bell's inequality.  To the extent that even this small result has been long thought impossible, realism has new cause for hope.

\section*{Acknowledgments}
The author would like to acknowledge conceptual inspiration from H. Price and R. Feynman, and technical inspiration from L. Schulman.  R.A. Linck and C.H. Salazar-Lazaro provided crucial assistance in developing the toy Lagrangian, and additional improvements arose thanks to R. Sutherland, W.M. Stuckey, J. Finkelstein, and H. Price.


\begin{thebibliography}{}
\bibitem{footnote0}
This definition is meant to exclude deBroglie-Bohm approaches, where the guiding field resides in multiparticle configuration space rather than spacetime.
\bibitem{footnote1}
The path to the Schr\"odinger equation is more convoluted; a scalar field Lagrangian generates the Klein-Gordon equation, whose parameter space needs to be halved as in section III.B before taking the non-relativistic limit; see \cite{KGWE}.
\end{thebibliography}
\end{document}